\documentclass[12pt]{article}
\usepackage{a4wide,epsf}

\begin{document}
 
\begin{titlepage}
\renewcommand{\thefootnote}{\fnsymbol{footnote}}
\makebox[2cm]{}\\[-1in]
\begin{flushright}
\begin{tabular}{l}
hep-ph/9605233\\
May 1996
\end{tabular}
\end{flushright}
\vskip0.4cm
\begin{center}
{\Large\bf The Exclusive Decay $B\to\rho e \nu$ Beyond\\[4pt]
  Model Calculations
}
 
\vspace{2cm}
 
Patricia Ball
 
\vspace{0.7cm}
 
{\em CERN, Theory Division, CH--1211 Gen\`{e}ve 23, Switzerland}
 
\vfill
 
{\bf Abstract\\[5pt]}
\parbox[t]{\textwidth}{
Due to its comparatively theoretical ``simplicity'', the decay channel
$B\to\rho e \nu$ offers one of the best possibilities to determine
the CKM matrix element $|V_{ub}|$ accurately. I present a new
calculation of the relevant hadronic form factors from light-cone sum
rules. I also review the results from lattice calculations and find
that they agree with the results from light-cone sum rules where
comparison is possible.\\
This paper relies on work done in collaboration with V.M.\ Braun.
}
 
\vspace*{5cm}
 
{\em
to appear in the Proceedings of the XXXIth Rencontres de Moriond,\\
``Electroweak Interactions'', Les Arcs, France, March 1996
}
\end{center}
\end{titlepage}
\renewcommand{\thefootnote}{\arabic{footnote}}
\setcounter{footnote}{0}
\newpage
\begin{figure}[h]
$$
\epsffile{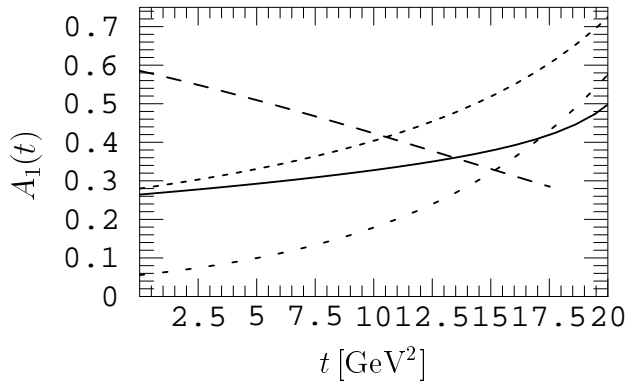}
$$
\vspace*{-1.2cm}
\caption[]{A sample of predictions for $A_1(t)$ from quark models and QCD
  sum rules. Solid line: light-cone sum rules \protect{\cite{BB}}; long
  dashes: three-point sum rules \protect{\cite{PRD48}}; short dashes
  with short spaces: BWS model \protect{\cite{BWS}}; short dashes with long
  spaces: ISGW model \protect{\cite{ISGW}}.}
\end{figure}
With the increasing statistics of experimental data from CLEO, the
determination of $|V_{ub}|$ from the decay channel $B\to\rho e \nu$
becomes more and more feasible. Experimentally, $b\to u$ transitions
are visible only above the kinematical threshold for charm production,
i.e.\ for electron energies $E_e>2.3\,$GeV in semileptonic decays. The
theoretical description of inclusive decays being exceedingly
difficult in that region, it seems more promising to look into the
exclusive channels. Here it is naturally the $B\to\pi$ and $B\to\rho$
transitions that, if at all, are tractable theoretically, and of these
the latter is the most prospective one, as it is expected to be
strongly peaked in the observable region, cf.\ Fig.~11 in Ref.~\cite{PRD48}.
It is thus timely to review the existing theoretical predictions for
these decays.

Let us begin with definig the relevant hadronic matrix element for
$B\to\rho$ transitions:
\begin{eqnarray}
\langle \rho,\lambda | (V-A)_\mu | B \rangle & = &
-i (m_B + m_\rho) A_1(t) \epsilon_\mu^{*(\lambda)} +
\frac{iA_2(t)}{m_B
+ m_\rho} (\epsilon^{*(\lambda)}p_B) (p_B+p_\rho)_\mu\nonumber\\
& & {} + \frac{iA_3(t)}{m_B + m_\rho} (\epsilon^{*(\lambda)}p_B)
(p_B-p_\rho)_\mu + \frac{2V(t)}{m_B + m_\rho}
\epsilon_\mu^{\phantom{\mu}\nu\rho\sigma}\epsilon_\nu^{*(\lambda)}
p_{B\rho} p_{\rho\sigma},\makebox[0.8cm]{}\label{eq:ME}
\end{eqnarray}
where the four form factors $A_{1,2,3}$ and $V$ depend on the momentum
transfer $t=(p_B-p_\rho)^2$ to the leptons; in the limit of vanishing
lepton mass $A_3$ does not contribute to the semileptonic decay rate
and will not be considered in this note. $\lambda$ is the
polarization of the $\rho$ meson. 

Fig.~1 shows a sample of predictions for
the form factor $A_1$, the situation with the others is similar. It is
obvious that the spread of predictions prevents any reliable
prediction of the decay rate. The
problem is essentially the large range of physical values of $t$,
$0\leq t \leq 20.3\,$GeV$^2$, which are all relevant for the
observable electron
spectrum. The lack of predictiveness of the dynamical properties of
form factors is a common feature of most quark models
\cite{BWS,ISGW,uebrigerquark} and forces them
to rely on ad hoc assumptions on the $t$-dependence like the pole
dominance hypothesis. However, for B decays an accurate {\em
  prediction} of the $t$-dependence is absolutely mandatory. At
present there are only two methods that can claim some right to make
such predictions founded in QCD, that is the QCD sum rules method on
the one hand and lattice calculations on the other hand. 

Let us shortly review the status of the latter ones. Due to the
restricted size of presently tractable lattices, typical masses of
simulated hadrons are around 1.5 to 3~GeV. Extraction of B physics
from the lattice thus requires the extrapolation of the results
obtained for small quark masses to the B scale. For form factors the
situation is even more complicated due to the presence of a second
potentially large scale, the momentum transfer $t$. The extrapolation
in the heavy mass $m_H$ can be done either ``na\"{\i}vely'', by
fitting to the ansatz $F=A+B/m_H$, where F stands for a generic form
factor \cite{abada,APE} (the extrapolation is done at fixed
three-momentum of the final
state meson in the rest system of the decaying particle, i.e.\ 
$t$ scales with the heavy mass, too),  or by using the guidance provided by heavy quark
effective theory, according to which near maximum $t$ the form factors
behave apart from logarithmic corrections in the heavy mass as \cite{HL}:
\begin{equation}
A_1(t_\mathrm{max}) \sim 1/\sqrt{m_H},\quad A_2(t_\mathrm{max}) \sim 
V(t_\mathrm{max}) \sim \sqrt{m_H}.\label{eq:scaling}
\end{equation}
The quantity constant in the heavy quark limit, i.e.\ $Fm_H^{\pm
  1/2}$, is then fitted by a polynomial in the inverse heavy mass,
cf.\ \cite{APE,UKQCD}. Working near $t_\mathrm{max}$ poses however
certain problems: first it is not clear in which range of $t$
the above scaling laws remain valid. Second, one is clearly interested
in the form factors in the full range of $t$, or, as for $B\to
K^*\gamma$, only in the value at $t=0$. Thus, without knowledge of the
functional $t$-dependence of the form factors the extrapolation to
smaller values of $t$ becomes
model-dependent\footnote{Actually most collaborations use a monopole
  behaviour, cf.\ \cite{abada,APE,UKQCD}, with which the
  available data at different (large) $t$ are consistent, without,
  however, being
  conclusive.}. For $B\to\pi$ transitions it was attempted to
restrict the functional $t$-dependence, given some values at large
$t$, from unitarity arguments \cite{unitarity}, but to my
knowledge no similar bounds for the $B\to\rho$ form factors exist to
date.

Another possibility advocated in Refs.~\cite{UKQCD,wuppertal} is to fix
$t=0$ and then to extrapolate the data in the heavy mass. As stressed in
\cite{wuppertal}, this method avoids the model-dependence from
extrapolating down to $t=0$, but it introduces another one, namely the
leading behaviour of the form factors at $t=0$ in the heavy mass,
which cannot be obtained from heavy quark effective theory.
Ref.~\cite{UKQCD} fits to $A_1(0)\sim m_H^{-3/2}$ (plus $1/m_H$
corrections), which relies on the scaling law (\ref{eq:scaling}) and
an assumed monopol pole behaviour of $A_1(t)$.
Ref.~\cite{wuppertal} tries three different fits with $F(0)\sim
m_H^{-1/2,0,1/2}$. Unfortunately, the data available to date do not
allow to distinguish between different powers of 
$m_H$, although the extrapolated values are rather
sensitive to it. In view of this difficulty, I would like to draw the
lattice community's attention to the fact that the leading behaviour
of form factors at $t=0$ in the heavy quark limit {\em can} be
extracted from QCD and turns out to be a decrease with $m_H^{3/2}$. 
This behaviour follows from the
asymptotic form of the leading twist $\rho$ meson light-cone wave 
functions in QCD. A comprehensive review on light-cone wave functions
can be found in Ref.~\cite{CZreport}, whereas the heavy quark mass
limit of form factors at $t=0$ was elaborated on in Refs.~\cite{CZ90,ABS,BB}.
I hope that in future calculations this behaviour will be
taken into account and lattice simulations will yield form factors
both at $t=0$ and large $t$, which would thus allow
to replace the hitherto used {\em extrapolations} in $t$ by {\em 
interpolations}.

Let me now turn to QCD sum rules. For the calculation of
heavy-to-light transition form factors there exist actually two types
of them. The more ``traditional'' ones are the ``three-point sum rules''
based on the classical approach of Shifman, Vainshtein and Zakharov 
\cite{SVZ} and rely on the
expansion of a three-point correlation function in terms of local
operators, whose vacuum matrix elements, the condensates, reflect the
complicated nature of the QCD vacuum and, following Ref.~\cite{SVZ},
 are used to describe
non-perturbative corrections to the perturbative three-point
functions. This  method was applied to $B\to\rho$
transitions in Refs.~\cite{schrott,PRD48}.

A modification of that classical approach are the ``light-cone sum
rules'' \cite{LC,CZ90,BBD} that combine the QCD sum rule technique
with an expansion in terms of (non-local) operators of definite {\em
  twist} (instead of dimension), whose matrix elements over the vacuum
and the light hadron are described by light-cone wave functions.
Predictions for $B\to\rho$
form factors from this approach were given in Refs.~\cite{ABS,BB}.

As discussed in Ref.~\cite{BB}, reliable predictions of form factors
from three-point sum rules are restricted to large values of $t$, but
fail for small $t$, i.e.\ $t\sim O(m_b)$. At
small $t$ the condensate contributions to these sum rules diverge in
the heavy quark limit (for instance $A_1(0)$ blows up as $m_b^{1/2}$),
which is in contradiction to what one would expect intuitively
and also different from the
behaviour of the perturbative contributions to the sum
rules. Technically, three-point sum rules fail to describe
 large transverse distances in the hadronization of the $\rho$ meson,
 which, however, are essential to one of the two dominant
 hadronization processses, the Feynman mechanism, cf.\ \cite{CZ90,BB}. These
 contributions are taken into account correctly in the light-cone sum rules
 approach, whose central objects are matrix elements of non-local
 operators that are expressed in terms of light-cone wave functions.
 For the $\rho$ meson there are four wave functions up to twist three,
 which are of the generic type
\begin{equation}
\left.\langle 0 | \bar{u}(x)\Gamma d(0) | \rho(p)\rangle\right|_{x^2=0} \sim
\int\limits_0^1\!\! du \, e^{-iupx}\,\Phi(u).
\end{equation}
Here $\Gamma$ is some Dirac structure and $u$ is the momentum fraction of the
$\rho$ carried by the quark. The wave function $\Phi$ can be interpreted as the
probability amplitude to find the $\rho$ in a state with minimum
number of Fock constituents and at small transverse separation. The
twist two wave functions of the $\rho$ were reexamined recenctly in 
Ref.~\cite{rhoWF}.

The results for the form factors both from light-cone sum rules and
lattice calculations\footnote{I only give the results
  from the UKQCD collaboration \cite{UKQCD}, which are the only ones
  that do not rely on the pole dominance hypothesis.} are shown in
Fig.~2. The agreement of light-cone sum rules and lattice results
within the error-bars is striking, in particular for $A_1$, which dominates
the decay rate at large $t$. Comparing with the
pole-dominance hypothesis, I find that $A_1$ increases, but less than
a monopole, $A_2$ is comparable with a monopole, whereas $V$ increases
nearly as a dipole. As for the errors, there exists a very
sophisticated error-analysis from lattice, whereas a corresponding
analysis for the light-cone sum rule results it is not that easy. Neglecting 
possible systematic errors of the method, I obtain from varying the
input parameters within reasonable ranges, in particular $m_b = (4.8\pm
0.2)\,$GeV:
$$
\Delta A_1 =\pm 0.06,\quad \Delta A_2 = {}^{+0.09}_{-0.07},\quad
\Delta V = {}^{+0.10}_{-0.09},
$$
nearly independent of $t$. For the integrated rates I obtain
\begin{equation}
B(B\to\rho e \nu) = (21.8\pm 6.5)|V_{ub}|^2,\quad \Gamma_L/\Gamma_T =
0.52\pm 0.10,\quad\Gamma_+/\Gamma_- = 0.016\pm 0.003.
\end{equation}
In Fig.~3 I also show the spectra $d\Gamma/dE$ over the electron energy
and $d\Gamma/dt$. It is now up to my experimental colleagues to
provide better data that would allow to confirm or exclude the
predictions from light-cone sum rules and lattice calculations.
\begin{figure}
$$
\epsfxsize=\textwidth
\epsfbox{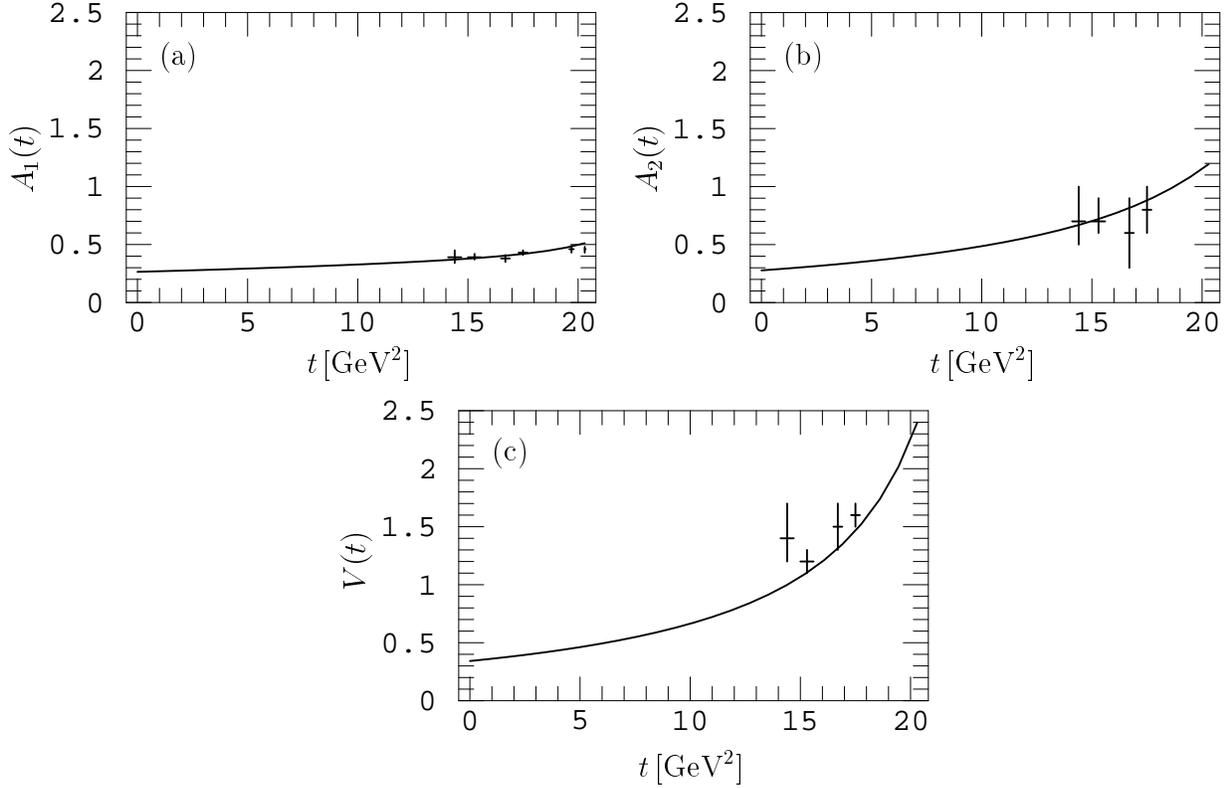}
$$
\vspace*{-1.2cm}
\caption[]{The semileptonic form factors as functions of $t$ from
  light-cone sum rules \cite{BB} (solid lines) and lattice
  calculations (UKQCD) \cite{UKQCD} (crosses).}
\end{figure}
\begin{figure}
$$
\epsfxsize=\textwidth
\epsfbox{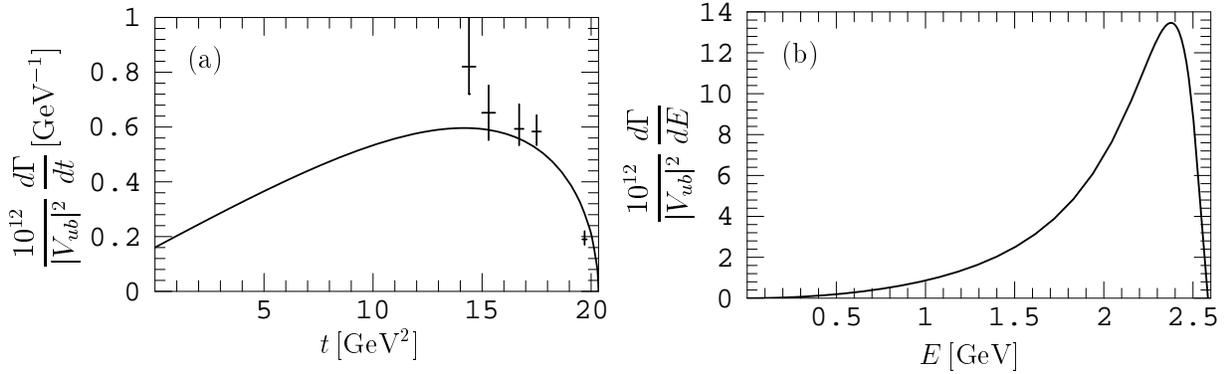}
$$
\vspace*{-1.2cm}
\caption[]{The spectrum $d\Gamma/dt$ for the decay $B\to\rho e \nu$
  from light-cone sum rules (solid line) and lattice calculations
  (crosses). (b) The electron spectrum from light-cone sum rules (not
  available from lattice due to restricted information in $t$).}
\end{figure}

\end{document}